\documentclass[useAMS]{mn2e}

\usepackage{epstopdf}
\usepackage{graphicx}

\def\simlt{\lower.5ex\hbox{$\; \buildrel < \over \sim \;$}}
\def\simgt{\lower.5ex\hbox{$\; \buildrel > \over \sim \;$}}

\title[FUV variability in V3885 Sgr]{Orbital and stochastic
far-UV variability in the nova-like system V3885 Sgr}
\author[R.K. Prinja et al.]{R.K. Prinja$^{1}$\thanks{E-mail: 
rkp@star.ucl.ac.uk},
K.S. Long$^{2}$,
M.T. Richards$^{3}$,
D.K. Witherick$^{1}$,
L.W. Peck$^{1}$\\
$^{1}$Dept. of Physics {\&} Astronomy, University College London, Gower Street, London WC1E 6BT \\
$^{2}$Space Telescope Science Institute, 3700 San Martin Drive, Baltimore, MD 21218, USA \\
$^{3}$Department of Astronomy {\&} Astrophysics, Pennsylvania State University, 525 
Davey Laboratory, University Park, PA 16802, USA\\
}
\begin{document}

\date{Accepted 2010. Received 2010; in original form 2010}

\pagerange{\pageref{firstpage}--\pageref{lastpage}} \pubyear{2007}

\maketitle

\label{firstpage}

\begin{abstract}
Highly time-resolved time-tagged {\it FUSE} satellite spectroscopic data
are analysed to establish the far-ultraviolet (FUV) absorption line characteristics
of the nova-like cataclysmic variable binary, V3885 Sgr.
We determine the temporal behaviour of low (Ly$\beta$, C{\sc iii}, N{\sc iii})
and high (S{\sc iv}, P{\sc v}, O{\sc vi}) ion species, and
highlight {\it corresponding} orbital phase modulated changes
in these lines. On average the absorption troughs are blueshifted
due to a low velocity disc wind outflow.
Very rapid ($\sim$ 5 min) fluctuations in the absorption lines
are isolated, which are indicative of stochastic density changes.
Doppler tomograms of the FUV lines are calculated which provide
evidence for structures where a gas stream interacts with
the accretion disc.
We conclude that the line depth and velocity changes as
a function of orbital phase are consistent with an asymmetry
that has its origin in a line-emitting, localised disc-stream interaction region.
\end{abstract}


\section{Introduction}
Cataclysmic variables (CVs) are mass-exchanging binary stars in which
a white dwarf primary accretes material from a late main-sequence secondary.
The systems
represent an important stage in the life cycle of binary stars
undergoing common envelope evolution.
Of specific relevance to the investigations presented in this paper is the
fact that CVs also provide the setting for some of the most
observationally accessible accretion disc-driven outflows
in the Universe, since the binary systems are relatively bright
and have short ($\sim$ hours) orbital periods.
Outflows in the form of winds with velocities of $\simgt$1000 km s$^{-1}$
are evident in almost all high accretion rate CVs, and particularly
nova-like and dwarf novae in outburst.

Empirical properties of the high-speed, high-ionization CV winds have
been established very effectively from UV line transitions accessible
to the {\it IUE}, {\it HST} and {\it FUSE} space borne observatories
(see e.g Heap et al. 1978; Cardova {\&} Mason 1982;
Hassall et al. 1983; Knigge, Woods {\&} Drew 1995).
The observational transition between blueshifted absorption and
P~Cygni profiles in low inclination (face-on) CVs to broad
emission dominated UV lines in more edge-on systems has been
interpreted as evidence for the bipolarity of the outflows
(e.g. Drew 1987).
It is clear however that the morphologies of the UV lines
cannot be assumed to be independent of orbital phase or temporal
changes. Time-series {\it HST} UV and {\it FUSE} FUV spectroscopic
datasets have revealed extensive variability in the
UV resonance lines on $\sim$ hourly time-scales
(e.g. Baptista et al. 1995;
Prinja et al. 2000; Hartley et al. 2002; Prinja et al. 2003;
Froning 2005).
The (F)UV time-series studies have revealed an enigmatic variety
of characteristics. Several low inclination systems exhibit orbitally
modulated, blue-shifted wind features for which the origin of the
departure in axisymmetry remains unclear. A few nova-like CVs reveal
erratic episodic features that accelerate rapidly across the
absorption line profiles; others in contrast have mostly steady
wind-formed UV lines. Additional uncertainties concern the origin
of thermal emission that produces substantial emission components
in CVs viewed at low inclination.

\begin{figure*}
 \includegraphics[scale=0.8]{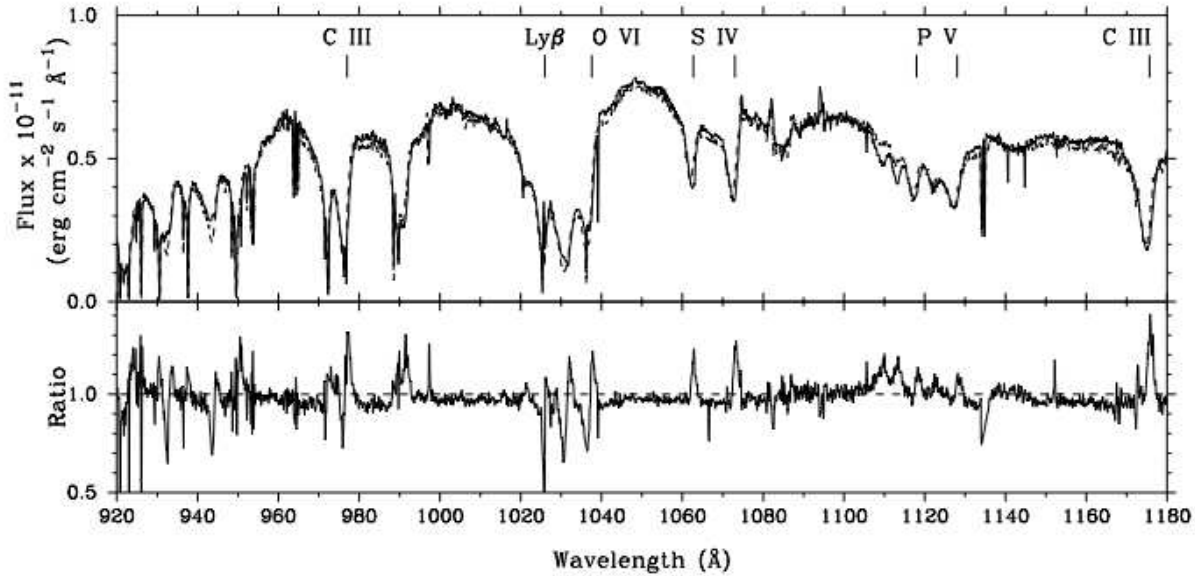}
 \caption{Mean {\it FUSE} spectra from Obs1 (2000 May 24;
black) and Obs2 (2003 Sept. 21/22; red, dashed).
The ratio of the two mean spectra from 2000 and 2003
is shown in the lower panel to illustrate the range of spectral
lines that vary in time.
}
\end{figure*}

The origin and driving of fast outflows in high-state CVs also
remains uncertain, primarily as the effective radiation force
in low (e.g. Proga, Stone {\&} Drew, 1997).
Though magnetic fields and thermal expansion have been considered
as mechanisms for potentially driving the disc winds, line-driven models
are currently the more fully developed in the context of bipolar CV
winds, and in the hydrodynamic and magneto-hydrodynamic limits
(see e.g. Proga 2005).
Proga, Stone {\&} Drew (1997) also highlighted the potentially
unsteady nature of the line-driven disc winds, with the consequent
rise due to instabilities of (stochastic) rapidly evolving (over several
minutes) clumps of gas; however in contrast the hydrodynamical models of
Pereyra, Kallman {\&} Blondin (1997) indicate more
stable outflows. Theoretical predictions 
of unsteady disc outflows are interesting in the context of
the rapid {\it FUSE} FUV spectroscopy presented here.

\subsection{Nova-like target V3885 Sgr}
We present here a study of the nova-like system V3885 Sgr, based on
highly time-resolved {\it FUSE} datasets. We aim to investigate
systematic changes in the UV lines and bipolar disc wind on
orbital and rapid ($\sim$ few minutes) time-scales. We examine whether
orbital-related changes are present and examine causal connections to an 
axi-symmetry of
the entire bipolar outflow. The
alternative of a non-wind departure of symmetry due to
line-emitting hots spots or disc residing structures is also
investigated.

V3885 Sgr is a bright, non-eclipsing CV (e.g. Warner 1995).
The recent studies of Ribeiro {\&} Diaz (2007) and Linnell et al. (2009)
have provided firm values for the orbital, disc and binary parameters.
Some adopted values are listed in Table 1.


\begin{table}
 \centering
\caption{V3885 Sgr adopted system parameters.}
  \begin{tabular}{lll}
  \hline
Parameter & Value & Reference  \\
\hline

Orbital Period & 0.20716 days & Ribeiro {\&} Diaz (2007) \\
Inclination & 65$^o$ $\pm$ 2$^o$  & Linnell et al. (2009) \\
Systemic velocity & $-$45 km s$^{-1}$ & Ribeiro {\&} Diaz (2007) \\
\hline

\end{tabular}
\end{table}


Linnell et al. also provide an extensive line list for the
{\it FUSE} spectrum of V3885 Sgr. The disc wind in V3885 Sgr
was previously studied by Hartley et al. (2002), using {\it HST}
STIS data to provide some evidence for rapid ($\sim$ 100 sec)
variations in the N{\sc v}, Si{\sc iv} and C{\sc iv} P Cygni lines;
their dataset was not extensive enough to examine orbitally
modulated behaviour.

\begin{figure}
 \includegraphics[scale=0.4]{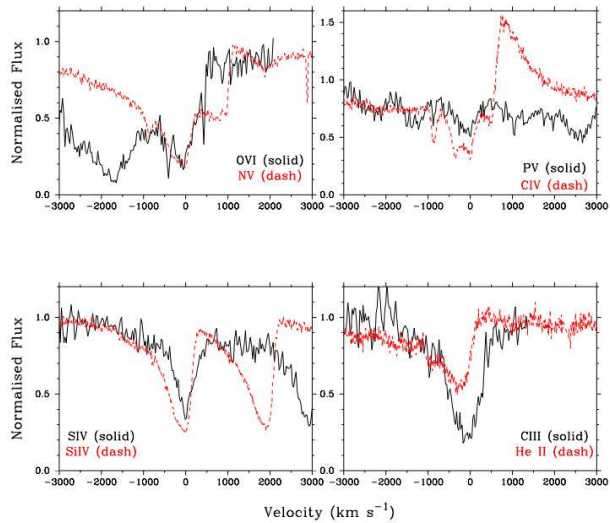}
 \caption{FUV (black) and UV (red; dashed) line profiles in 
individual {\it FUSE} and {\it HST} spectra of V3885 Sgr
for a range of high to low ion species.}
\end{figure}


\begin{figure*}
 \includegraphics[scale=0.85]{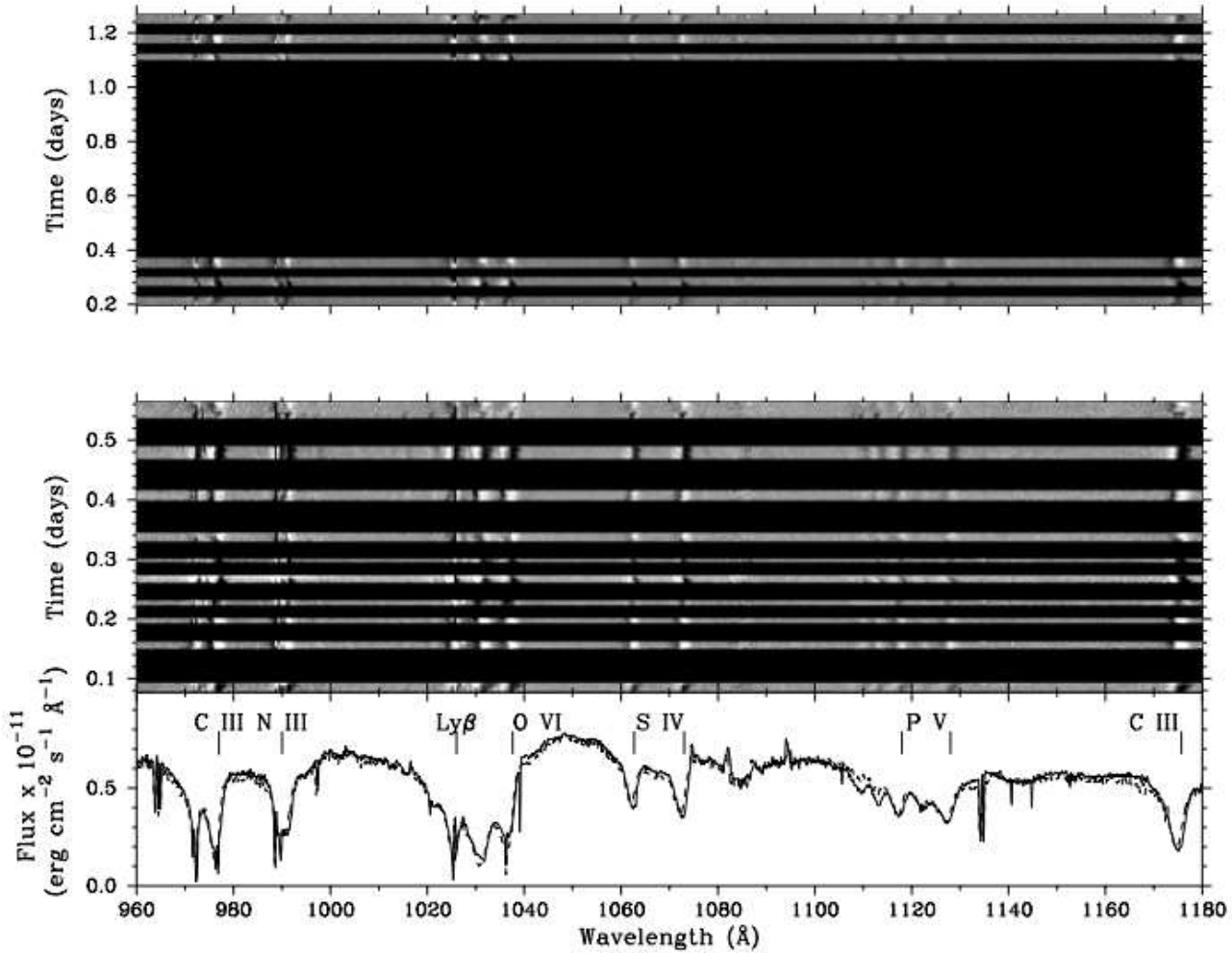}
 \caption{Grey-scale representation of variability across
the full TTAG {\it FUSE} spectral range in V3885 Sgr. The lower
and upper panels display the
the residual with respect to the mean of all spectra
for Obs1 (2000 May 24) and Obs2 (2003 Sept. 21/22), 
respectively. In all the grey-scales shown here the black and
white dynamic range levels are set to 0.6 and 1.3, respectively.
The bottom panel show the mean spectrum for the Obs1 (solid) and Obs2
(dashed) datasets.
}
\end{figure*}


\begin{figure}
 \includegraphics[scale=0.53]{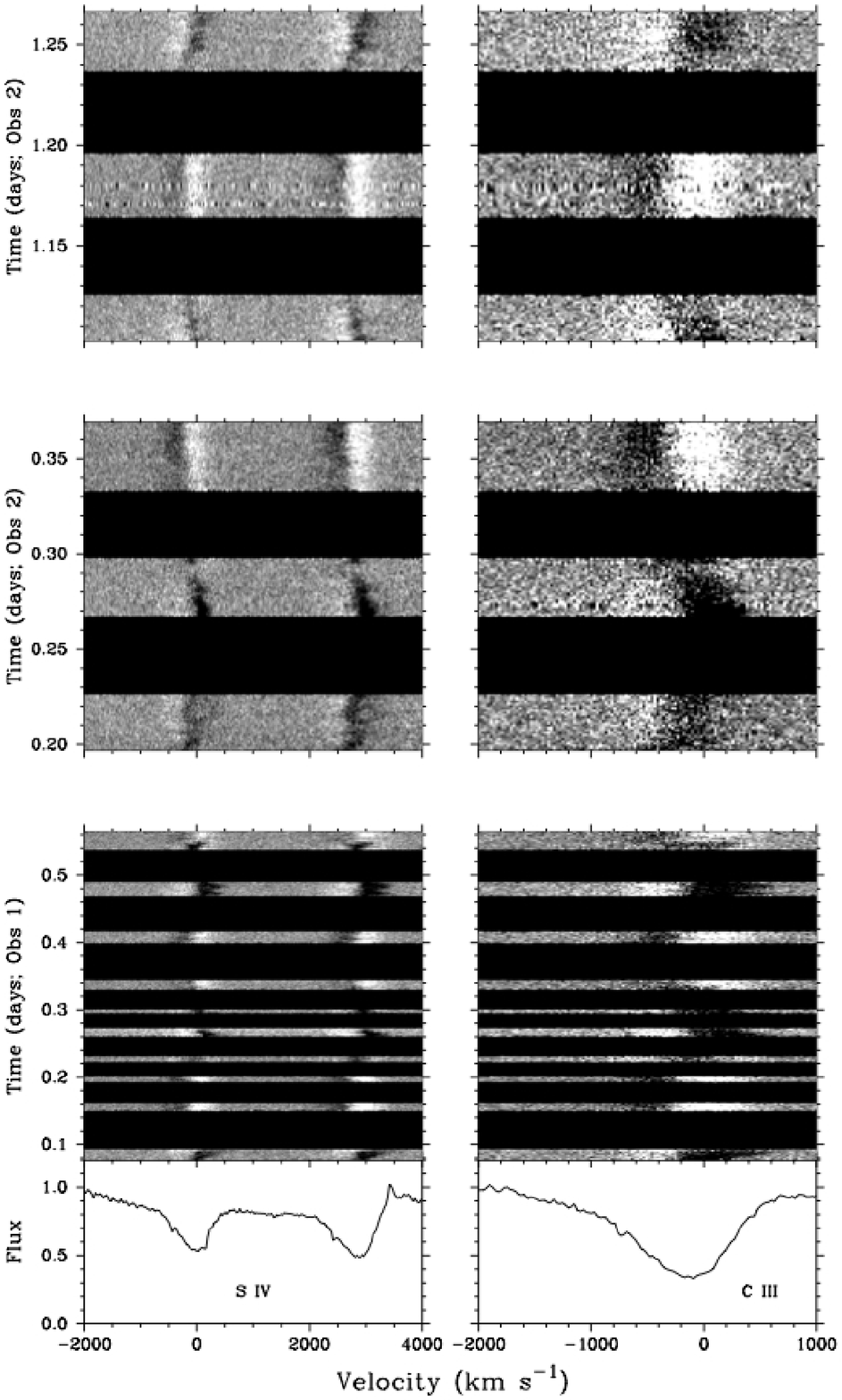}
\caption{Close-up of the time-variability in
S{\sc iv} $\lambda\lambda$1063, 1073 and
C{\sc iii} $\lambda$1175.67. The data have been
corrected for an adopted systemic (white dwarf) velocity
of $-$45 kms s$^{-1}$ (see Table 1)
The bottom panel show the mean spectrum for the Obs1
dataset.
}.
\end{figure}


\section{{\it FUSE} time-tagged dataset}

The {\it FUSE} satellite instruments are described in detail by
Moos et al. (2000).
Four spectra (SiC1, SiC2, LiF1, LiF2) are imaged in two segments
(A and B). The total of eight segments covers a spectral range
from $\sim$ 905{\AA} to 1187{\AA}, with sufficient overlap between
individual segments.
Our observations of V3885 Sgr were obtained in time-tag (TTAG) mode,
so that the arrival times of photons are recorded. The data 
were secured through the large aperture (30$^"$ $\times$ 30$^"$) and
then extracted in 100 sec bins.
The original exposure times of the three
raw exposures re-reduced for this project were 12,900 sec
(Prog. ID P197; Malina; 2000 May 24) and
8300 sec and 71000 sec (Prog. ID D905; Froning; 2003 September 21/22).
After excluding a few very noisy spectra, the final dataset analysed
comprises 126 {\it FUSE} spectra spanning $\sim$ 11.7 hours
on 2000 May 24 (`Obs1') and 155 spectra over $\sim$ 25.7 hours
between 2003 September 21/22 (`Obs2').
The time-sequence
in Obs 2 is interrupted by a data gap of $\sim$ 17.6 hours.
The spectral resolution of an individual spectrum
is $\sim$ 0.1 {\AA}, with continuum signal-to-noise $\sim$ 10.

The mean spectrum for Obs1 and Obs2 is shown in Fig. 1.
A familiar set of high-state CV lines are seen.
The lower plot is Fig. 1 shows the ratio of the two mean profiles
from 2000 and 2003, and demonstrates that 
line profile variability is extensive across the {\it FUSE} range.
In contrast the continuum
flux levels agree well over the almost 3 year time-span 
between.

Close-up views of key spectral lines are shown in Fig. 2, where
line profiles from a representative individual {\it FUSE}
spectrum are show, together with UV profiles
from {\it HST} STIS (Prog. i.d. O5BI06010; 2000 November 13;
described in detail by Hartley et al. 2002).
Additional individual spectra are also shown later in Fig. 9.
The maximum blueward velocity of the lines does {\it not}
differ significantly between the high ions (O{\sc iv}, N{\sc v})
and low ion species (He{\sc ii}, C{\sc iii}). (Note that
from herein the data have been corrected for a
systemic (white dwarf) velocity of $-$45 km s$^{-1}$; see Table 1).
O{\sc iv}, N{\sc v}, S{\sc iv}, Si{\sc iv}, C{\sc iii} all exhibit
some blueward asymmetry. One of our goals here is to assess
what component of the variable total absorption is non-outflow, i.e.
due to accretion disc-stream interaction instead. 
The individual {\it FUSE} spectra from 2000 and 2003 do not
exhibit any cases of strong emission components with respect
to the local continuum.
The C{\sc IV} $\lambda\lambda$1550
resonance line is undoubtedly  wind-formed and is the only
spectral line in V3885 Sgr to reveal an unambiguous emission feature
(see also Hartley et al. 2002).

\section{Variability characteristics}

The {\it FUSE} data sets of V3885 Sgr assembled here permit us
to explore variability across a wide
range of spectral lines, covering low and high ions.
To globally compare the temporal behaviour of the lines, dynamic
spectra (grey-scale images) were constructed 
covering the entire TTAG datasets for
Obs1 and Obs2. The respective images are shown in Fig. 3, as a function
of time. To enhance the contrast in the relative profile
changes, quotients are shown between the individual TTAG spectra
and the respective (Obs1 or Obs2) overall mean spectrum.
As  was the case in our previous studies (e.g. Prinja et al. 2003, RW Sex;
Prinja et al. 2004, V592 Cas),
almost every reasonably developed absorption line in V3885 Sgr is varying,
and the changes are in concert between low and high ionization
species. The `twisted' black
and white tracks seen in the images in Fig. 3 correspond well between
the different lines, and are
indicative of an `S-wave' type velocity motion of the absorption lines
(see Sect 3.1).

The time-variability evident in S{\sc iv} $\lambda\lambda$1063, 1073
and C{\sc iii} $\lambda$1176 is shown in closer detail in Fig. 4, this
time as a function of velocity. The behaviour agrees very well between
the two lines. Systematic changes are seen {\it both} blueward
and redward of the rest velocity, but with a greater blueward
extension to $\sim$ $-$500 km s$^{-1}$. The motion demonstrated here
over the $\sim$ hourly time-scales is a velocity swaying of the entire
absorption profiles, as opposed to changes confined to
the line wings
or localised optical depth structure moving across the troughs.

To estimate the significance of the line profile variability as a
function of velocity, we applied the temporal variance spectrum (TVS)
analyses (see e.g. e.g. Fullerton et al. 1996), and
computed for each spectral line:
\begin{equation}
({\rm
TVS})_i=\sigma_0^2\frac{1}{N-1}\sum_{j=1}^N\left(\frac{S_{ij}-\bar{S}_i}
                {\sigma_{jc}\sqrt{S_{ij}}}\right)^2~,
\end{equation}
where
$S_{ij}$ is the normalized intensity of the $i$th pixel in the $j$th
spectrum,
      $\bar{S}_i$ is the weighted mean of the normalized intensity,
      $\sigma_{jc}$ is the inverse of $S/N$ of spectrum $j$ measured in
                        an adjacent continuum band, and
      $\sigma_0^2 =
\left[\frac{1}{N}\sum_{j=1}^N\sigma_{jc}^{-2}\right]^{-1}$. The results
are shown as root mean square percentages (= TVS$^{1\over{2}}$ $\times$
100) in Fig. 5.
The RMS changes agree very well in structure, amplitude and velocity range
between O{\sc vi}, S{\sc iv} and C{\sc iii}. There are
two distinct components to the variability;
(i) variance almost symmetric about rest velocity, and
extending to $\pm$ 400 km s$^{-1}$, though perhaps more redward
than this range in C{\sc iii}.
We attribute this component of the TVS to the
corresponding velocity movement of the the lines, which
provides the dominant S-wave pattern seen in Fig. 4.
(ii) A weaker more blueward variance between $\sim$ $-$400 km s$^{-1}$
to $-$700 km s$^{-1}$.
Note that the absorptive line features in the FUV spectrum of
V3885 Sgr have a greater redward extent than e.g. was the case
in V592 Cas, RW Sex, BZ Cam (see references in Sect. 1).

\begin{figure}
 \includegraphics[scale=0.39]{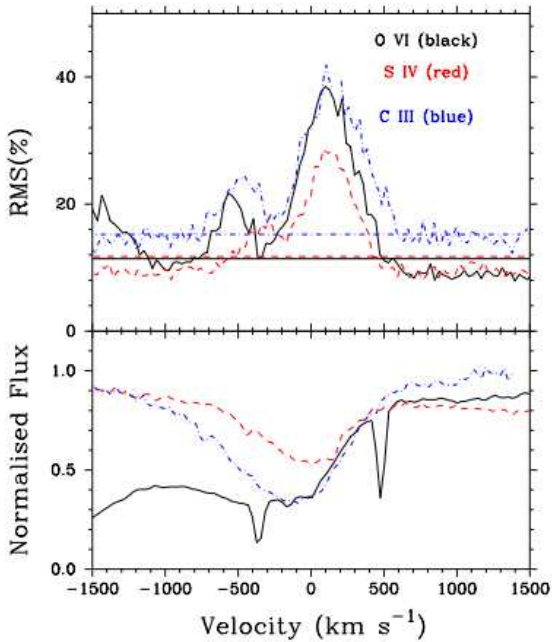}
 \caption{The variance (upper panel) and mean profiles (lower panel)
for O{\sc vi}, S{\sc iv} and C{\sc iii}.
The horizontal dashed lines indicate the 95{\%} confidence limit
for the variance.}
\end{figure}


\subsection{Orbital Phased-modulated behaviour}

The time-scales of variability evident in Figs. 3 and 4 are
commensurate with the generally accepted orbital period of
V3885 Sgr of $\sim$ 0.207 days (Table~1).
Our time-series dataset of V592 Cas is not as extended as
previous (optical) studies of this system that have established
orbital parameters, especially in the case of Obs2.
Nevertheless, we conducted
a periodogram analysis to parameterise the apparent cyclic behaviour.
The TTAG datasets for the S{\sc iv} $\lambda$1063 and
C{\sc iii} $\lambda$1176 lines were Fourier analysed
using the iterative CLEAN
algorithm (Roberts, Leh{\'a}r {\&} Dreher 1987) to
deconvolve the features of the
window function from the discrete Fourier transform.
(A gain of 0.5 with 100 iterations was used.)
The frequency range
sampled by the data from $\sim$ 0.5 days$^{-1}$ to
50 days$^{-1}$ was examined for periodic signals.
The only potentially significant peaks in the resulting power
spectra (see Fig. 6) are at
4.940 day$^{-1}$ (C{\sc iii}; Obs1),
5.206 day$^{-1}$ (S{\sc iv}; Obs1),
4.853 day$^{-1}$ (C{\sc iii}; Obs2), and
4.837 day$^{-1}$ (S{\sc iv}; Obs2).
We give greater weight to the more continuously sampled results
from Obs1, where the mean frequency corresponds to a period
of 0.197 days.
We estimate a $\sim$ 30{\%} uncertainty in these frequencies
based on the
half-width at half-maximum of the 
zeroth order peak of the window function Fourier transform.
The 0.197 day period is repeated over about 2.5 cycles
during the full span of the observations in Obs1.
We conclude that the major systematic hourly variations seen in the
FUV lines of V3885 Sgr are modulated on the orbital period of the
system.

\begin{figure}
 \includegraphics[scale=0.42]{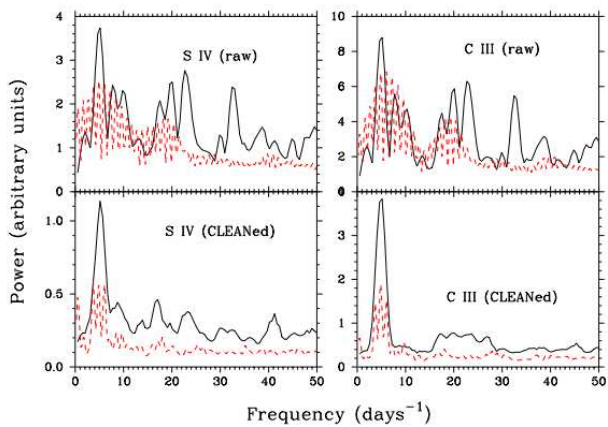}
 \caption{The raw (red; dotted) and CLEANed (black; solid)
Fourier power spectra for the  S{\sc iv} and C{\sc iii}
TTAG time-series data. The upper panels are for Obs2 (Sept. 2003)
and the lower panels for Obs1 (May 2000).}
\end{figure}


Grey-scale representations
of the individual quotient (with respect to the mean) line
profiles are shown in Fig. 7 phased on the generally accepted
orbital period (0.207 day; Table 1). The data for
S{\sc iv} and C{\sc iii} have been phased
using the accurate long-term ephemeris of Ribeiro {\&} Diaz (2007)
which is based on optical spectroscopy obtained between 1999
and 2002, i.e. straddling the epochs of our FUSE datasets; Sect. 2.
In Fig. 7, Phase = 0
corresponds to the +ve to $-$ve crossing in velocity of
radial velocity curves for the lines in Ribeiro {\&} Diaz (2007).
The cyclic modulation has a greater
blueward extent, but the changes are also clearly shallower
in optical depth on the blueward side,
between $\sim$ 0 and $-$ 300 km s$^{-1}$.
The strongest variations are close to or at rest velocity.

\begin{figure*}
 \includegraphics[scale=0.75]{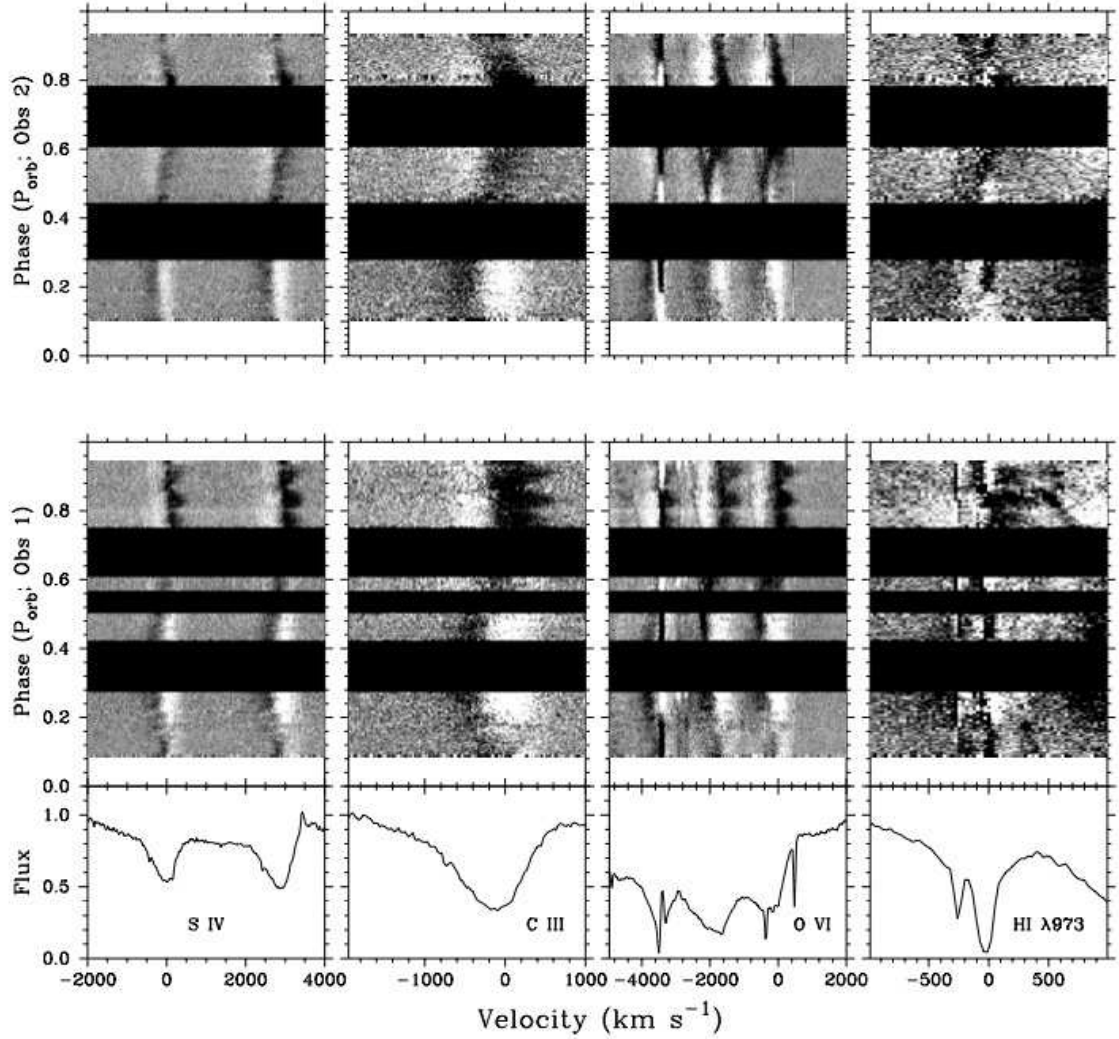}
 \caption{Grey-scale representations of variability in
S{\sc iv}, C{\sc iii}, O{\sc vi} and H{\sc i},
phased on the orbital period of
V3885 Sgr. Phase = 0 corresponds to the positive to negative
velocity crossing of the spectral lines for each dataset (Obs1
and Obs2).}
\end{figure*}


The line changes in V3885 Sgr were also measured by (least-squares) 
fitting the S{\sc iv} $\lambda$1062.66 component with single Gaussian 
model profiles, to record the central velocity, width and relative flux 
of the line. The blue component of this doublet is relatively
free of emission features and the central absorption trough
is well matched by the adopted Gaussian profile.
The central velocities determined from the
profile fits are plotted as a function of orbital
phase in Fig. 8; the velocity shift toward the blueward side
is about 100 km s$^{-1}$ greater than the redward motion (left-panel).
Though there is some scatter, the Gaussian fits also suggest
that the strength of the FUV absorption lines is greater
at redward velocities.
the line shifts toward blueward
velocities (right-hand panel).
The line-of-sight velocity dispersion is also variable,
as indicated by
the full-width at half-maximum of the fitted profile
in S{\sc iv}, which vary between $\sim$ 200 km s$^{-1}$ to
$\sim$ 550 km s$^{-1}$; there is however no obvious trend with
orbital phase.

\begin{figure*}
 \includegraphics[scale=0.85]{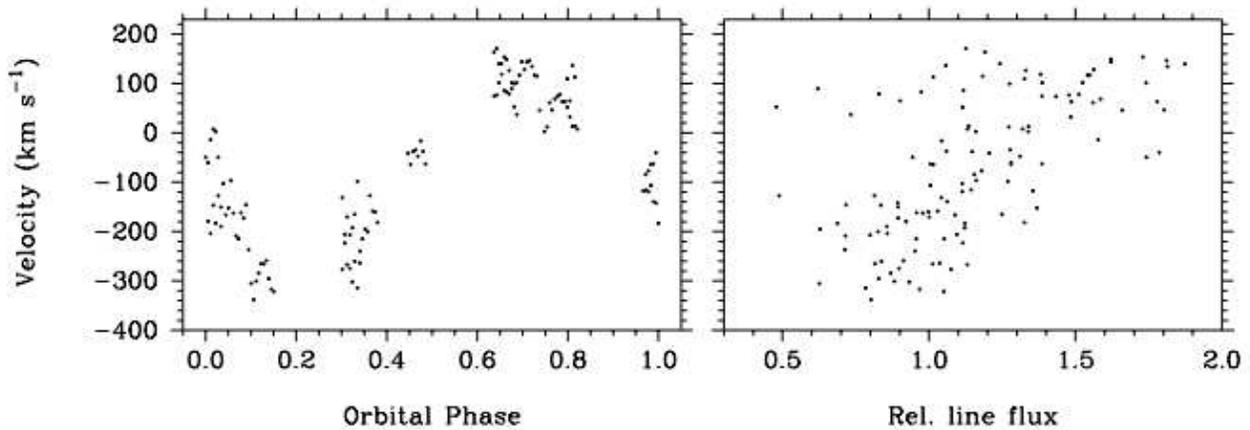}
 \caption{Left-hand panel $-$ The Gaussian-fitted central velocity
of S{\sc iv} $\lambda$1062.66 is plotted versus orbital phase.
Right-hand panel $-$ The central velocity is
plotted against relative absorption line strength
(in arbitrary units).
}
\end{figure*}


\subsection{Rapid variability}
Our {\it FUSE} dataset has established that the FUV resonance
and other absorption lines in V3885 Sgr vary cyclically on the
orbital period. The variation is characterised by velocity shifts
of several 100 km s$^{-1}$ and the optical depth changes are
stronger close to rest velocity than at blueward velocities.
One of the drivers for assembling {\it FUSE} TTAG datasets of V3885 Sgr
in 100 sec. bins is to search for evidence of very rapid ($\sim$ minutes)
line profile changes. Fluctuations on this short time-scale may betray
clumped gas arising from instabilities in the line-driven disc wind
(see e.g. the predictions of Proga, Stone {\&} Drew 1998), or perhaps
stochastic changes associated with other line absorbing structures such
as the accretion stream.
It is interesting to note that there is sub-structure
within the overall modulation. See for example the data between
phases $\sim$ 0.75 and 0.9 in Obs1 in Fig 7. The `jagged' structure
seen here is very well mimicked between the S{\sc iv}
doublets and also in C{\sc iii}.

Our approach was to first `subtract' the
primary orbital modulation of the lines and then examine
the residual spectra for very short time-scale changes
(i.e. less than $\sim$ 10{\%} of the orbital period).
The dataset in Obs1
was isolated into 10 different time-bins, that are defined
by continuous runs of data, each with no gaps greater than 0.03 days.
These continuous data bins are in fact seen in the lower
grey-scale images in Figs. 3 and 4, each separated by a black data gap.
The individual TTAG spectrum in each of these 10 bins was
then divided by the corresponding mean for the respective bins.
These mean profiles can be considered as
representative of the average line over $\sim$ 5{\%} to 7{\%}
of the orbital period. Figure 9 shows dynamic spectra for
the quotient
S{\sc iv} doublet lines taken
from 4 of the 10 continuous data bins, where additional
{\it very} rapid variability is clearly present.
These images, and the pair of line profiles shown below the
grey-scales, point to optical depth and velocity changes occurring
on time-scales down to 5 minutes or less. The upper-right panel
in Fig. 9, for example, shows two discrete `events' lasting
over about 7 minutes (i.e. $\sim$ 2{\%} of the orbital period).
These changes clearly do not follow the sinusoid of the
overall orbital motion (which we have in any case `removed'
to the 1st order). These short-timescale variations are
also evident in a correlated manner in high ion species
such as the O{\sc vi} doublet (see Fig. 7; Obs 1 data at phase
$\sim$ 0.8 to 0.82). The correspondence between different lines
suggests the rapid variability is due to density fluctuations
as opposed to changes in the ionization state.

\begin{figure}
 \includegraphics[scale=0.42]{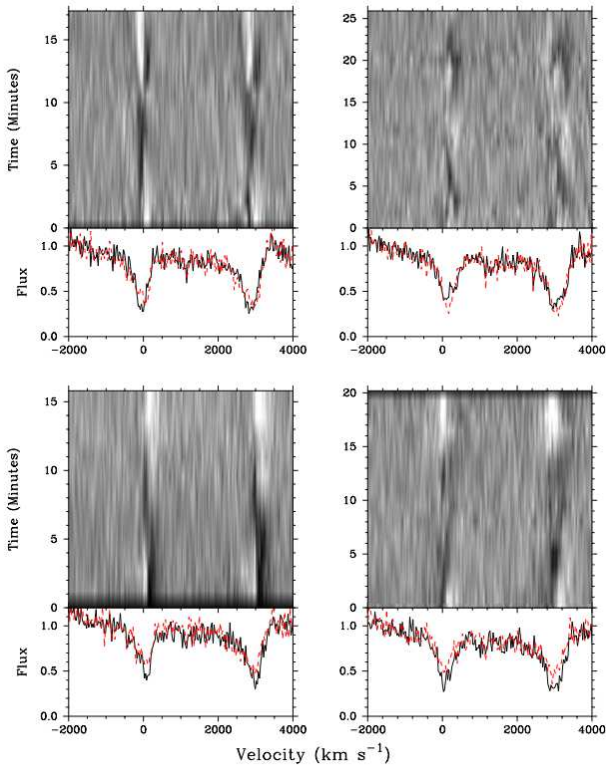}
 \caption{Dynamic spectra showing evidence for very rapid
fluctuations in the S{\sc iv} lines of V3885 Sgr. Changes
are evident on timescales down to 5 mins or less.
The lower panels below each grey-scale image show
pairs of spectra separated by just 3 minutes.}
\end{figure}


\begin{figure*}
 \includegraphics[scale=0.87]{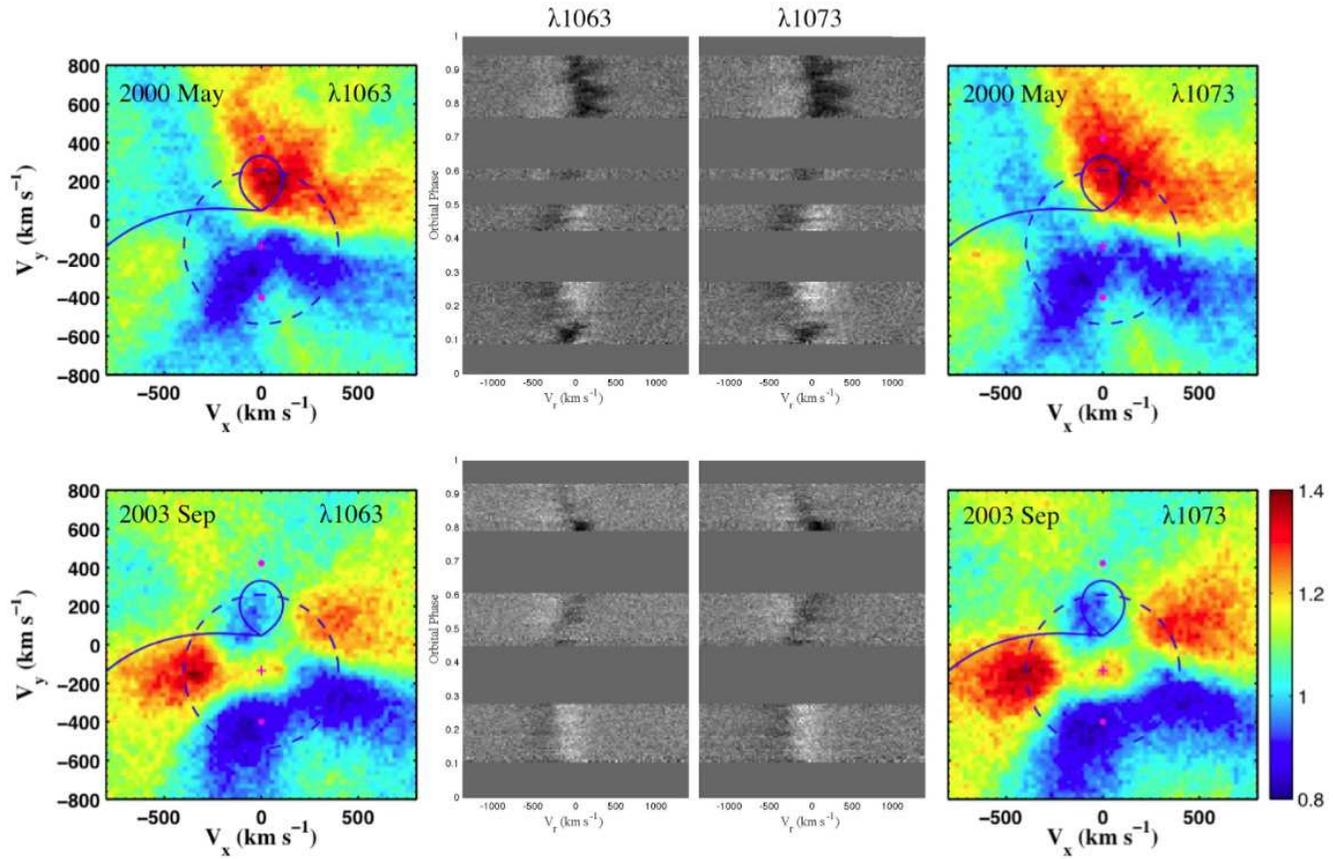}
 \caption{
2D Doppler tomograms based on difference profiles of the S{\sc iv}
doublet at two epochs (outer panels) with the corresponding trailed
spectrograms (inner panels). In the tomograms, the solid trajectory is
the gravitational free-fall path of the gas stream; the large dashed
circle marks the outer edge of a Keplerian disc; the plus sign marks the
velocity of the mass gainer, and the L2 and L3 points are represented by
the large dots along the line of centres.  The gaps in the velocity maps
correspond to the gaps in the phase coverage.
}
\end{figure*}


\section{Doppler tomography of S{\sc iv} absorption}

The image reconstruction method known as Doppler tomography has provided 
2D and 3D images of the velocity flows in a range of interacting binaries 
(e.g., Marsh {\&} Horne 1988; Richards, Albright {\&} Bowles 1995; 
Richards, Sharova and Agafonov 2010).  It has been applied to
CVs, X-ray and gamma-ray binaries, RS CVn 
binaries, and Algol-type binaries.  A diverse set of circumstellar 
structures has been identified in the tomograms including the 
gravitational stream of gas from the L1 point towards the mass gaining 
star, a transient or classical accretion disc, shock regions associated 
with interactions between the gas stream and the disc, an accretion 
annulus representing regions where the gas has slowed down after circling 
the mass gainer, and often emission sources indicative of magnetic 
phenomena such as prominences and coronal mass ejections.

Hartley et al. (2005) created Doppler tomograms of V3885 Sgr from
H$\gamma$, H$\beta$, and He{\sc i} $\lambda$4472
 spectra collected in 2003 Sep, and interpreted the results in terms 
of emission from
an accretion disc. They noted that the disc flow pattern could be 
explained either by
spiral shocks or by the tidally-thickened regions of the outer disc being 
illuminated by
light from the inner disc or boundary layer.
In addition, Ribiero and Diaz (2007) created
tomograms of V3885 Sgr based on H$\alpha$ and 
He{\sc i} $\lambda$6678 spectra collected from 1999 
Sep to 
2002 Jun.
The H$\alpha$ image was dominated by emission from the mass donor, while 
the He{\sc i} $\lambda$6678
image showed evidence of the accretion disc as well as emission from the 
mass donor.
These papers have shown that the tomography technique is a very
convenient way of looking at the complex line profile data in V3885.
The technique has been used in the case of Algol binaries to explore the
velocity regime in the cases of both optically thin and optically thick
lines (e.g., CX Dra, Richards et al. 2000; and TT Hya, Miller et al.
2007) with results that are consistent with the observed spectra.
Applying the tomography technique to optically thick line profiles
containing non-negligible emission components is challenging because of
the degeneracy between optical depth and emissivity embedded in the
Fourier slice theorem (also known as the Radon transform).  A careful
interpretation of the results is essential in these cases to demonstrate
the reliability of the tomography results.


\begin{table}
 \centering
\caption{V3885 Sgr adopted tomography parameters.}
  \begin{tabular}{ll}
  \hline
Parameter & Value \\
\hline

Orbital Period & 0.2071607 days \\
Systemic velocity & $-$45 km s$^{-1}$ \\
Mass ratio, $q$ & 0.679 \\
Primary velocity amplitude K$_1$ & 139 km s$^{-1}$ \\
Secondary velocity amplitude K$_2$ & 205 km s$^{-1}$ \\
Mass of primary & 0.7 M$_\odot$ \\
Radius of primary & 0.02 R$_\odot$ \\
\hline

\end{tabular}
\end{table}


We explore here whether tomography can offer any constraints on the line 
formation
regions of the variable FUV absorption in V3885 Sgr.
We add the caveat that the line formation of the FUV resonance lines
is likely to be complex due to contributions from multiple physical
sources such as the disc, disc$-$stream interaction, and the outflow.
Two dimensional images of V3885 Sgr were calculated using
a back-projection
tomography code (e.g. Richards 2004).
Since the observed
S{\sc iv} doublet 
profiles are
dominated by absorption, the tomograms were made from difference profiles 
by subtracting the mean spectrum for each epoch (Obs1, Obs2) 
from each respective observed spectrum. The coverage 
afforded by
our {\it FUSE} time-series is reasonably complete in phase except for 
phases 0.3$-$0.4 and 0.6$-$0.7
(e.g. Fig. 7), and high resolution spectra collected around the orbit of 
the binary are
required to produce a good image. Tomograms were calculated from
S{\sc iv} $\lambda\lambda$1063, 1073
spectra collected during observing runs on 2000 May 24 (Obs1) 
and 2003
Sep. 21/22 (Obs2). The parameters of V3885 Sgr adopted for 
the calculation
are listed in Table 2. The Doppler maps based on the
S{\sc iv} difference 
spectra are shown
in Fig. 10 and reveal the line emission regions in Doppler coordinates 
($V_x$,$V_y$), where
low velocity patterns correspond to the inner regions of the images, and 
high velocity
parts of the inner disc are mapped to the outer portions of the tomograms.

The tomograms reveal closely corresponding structures for the separate 
$\lambda$1063 (blue) and
$\lambda$1073 (red) components of the
S{\sc iv} doublet, although the 
absorption 
features in the blue
difference spectra were typically weaker than in the red spectra. There 
are similarities
between the tomograms derived at the two epochs, which is expected since 
the orbital
modulation is present at both epochs (e.g. Fig. 7). However, some clear 
differences are
evident. For example, the intense emission source seen in Obs1 near the 
velocity of the
mass donor (at $V_x$ $\sim$ 0 to 100 km s$^{-1}$
and $V_y$ $\sim$ 200 to 500 km s$^{-1}$) is 
significantly
reduced in Obs2 (see Fig. 10). During Obs2, the most intense emission 
arose from two
regions on opposing sides of the accretion disc around the white dwarf 
with $V_x$ velocities
of $-$250 to $-$700 km/s and +300 to 700 km/s. The blueshifted region is 
close 
to where the
gas stream would interact with the accretion disc, and the redshifted part 
can be viewed
half an orbit away, as emission from the 
stream-disk impact region
moving away from us.

\begin{figure*}
 \includegraphics[scale=0.87]{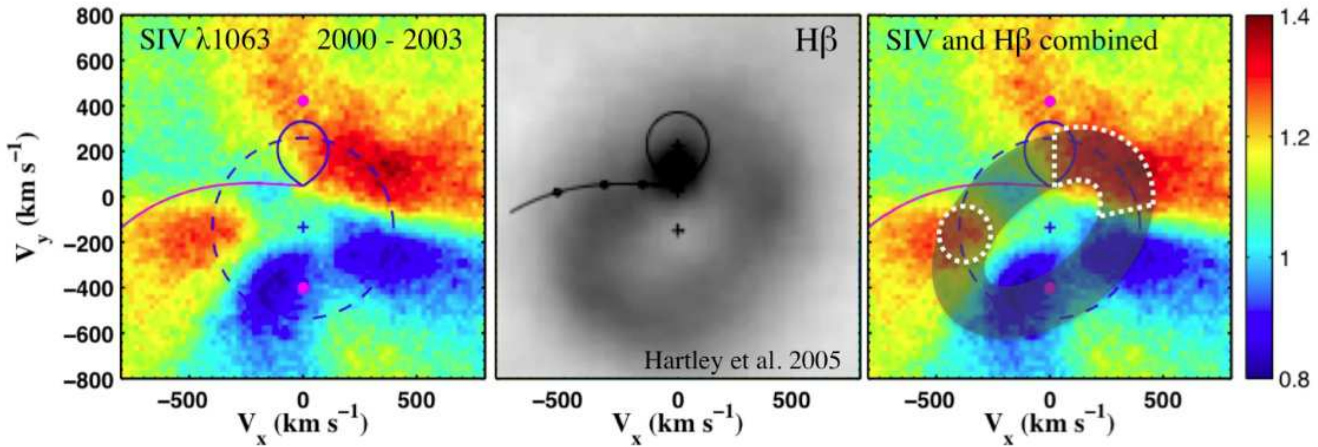}
 \caption{
2D Doppler tomogram based on the difference profiles of the
S{\sc iv} lambda 1063 line for both epochs (left frame) compared to the
Hartley et al. (2005) tomogram of the H$\beta$ line over the same velocity
range (middle frame), and the combined S{\sc iv} and H$\beta$ images.  The
overlapping regions are outlined by the dashed areas. The most intense
parts of the S{\sc iv} tomograms overlap with the brightest parts of the
accretion disc, so the S{\sc iv} results are consistent with the H$\beta$ images.
}
\end{figure*}


Fig. 11 shows the comparison between the S{\sc iv} tomograms  and the H$\beta$
tomograms from Hartley et al. (2005).  The main difference between the
images results from the incomplete phase coverage for the S{\sc iv} data.
However, the most intense parts of the S{\sc iv} tomograms overlap with the
brightest parts of the accretion disc identified in the H$\beta$ tomograms.
The regions of overlap are illustrated by the dashed areas on the image
in the right frame of Fig. 11.   The trailed spectrograms of the S{\sc iv}
lines display an S-wave pattern suggesting that a complete data set
might produce a more disk-like image as seen in the H$\beta$ tomogram.

 \section{Discussion}

We have analysed the FUV absorption line characteristics of the
nova-like CV V3885 Sgr. {\it FUSE} satellite spectra secured in TTAG
mode were extracted in 100 sec bins to provide intensive time-series
datasets comprising of 126 spectra in 2000 May and 155 spectra in
2003 September. {\it All} the well developed absorption lines in the
{\it FUSE} range are tightly modulated on the system orbital period
of $\sim$ 0.207 days. The TTAG data additionally reveal very rapid
($\sim$ 5 to 10 mins) fluctuations in the absorption lines, which
are indicative of stochastic density changes. We have calculated the
first FUV Doppler tomograms of V3885 Sgr and highlight a variable
line-emitting source where the gas stream from the secondary interacts
with the accretion disc. The FUV lines are on average blueshifted
thus suggesting a disc outflow is
also present (see e.g. Fig. 2 and {\it HST} data
of Hartley et al. 2002).

\begin{figure*}
 \includegraphics[scale=0.87]{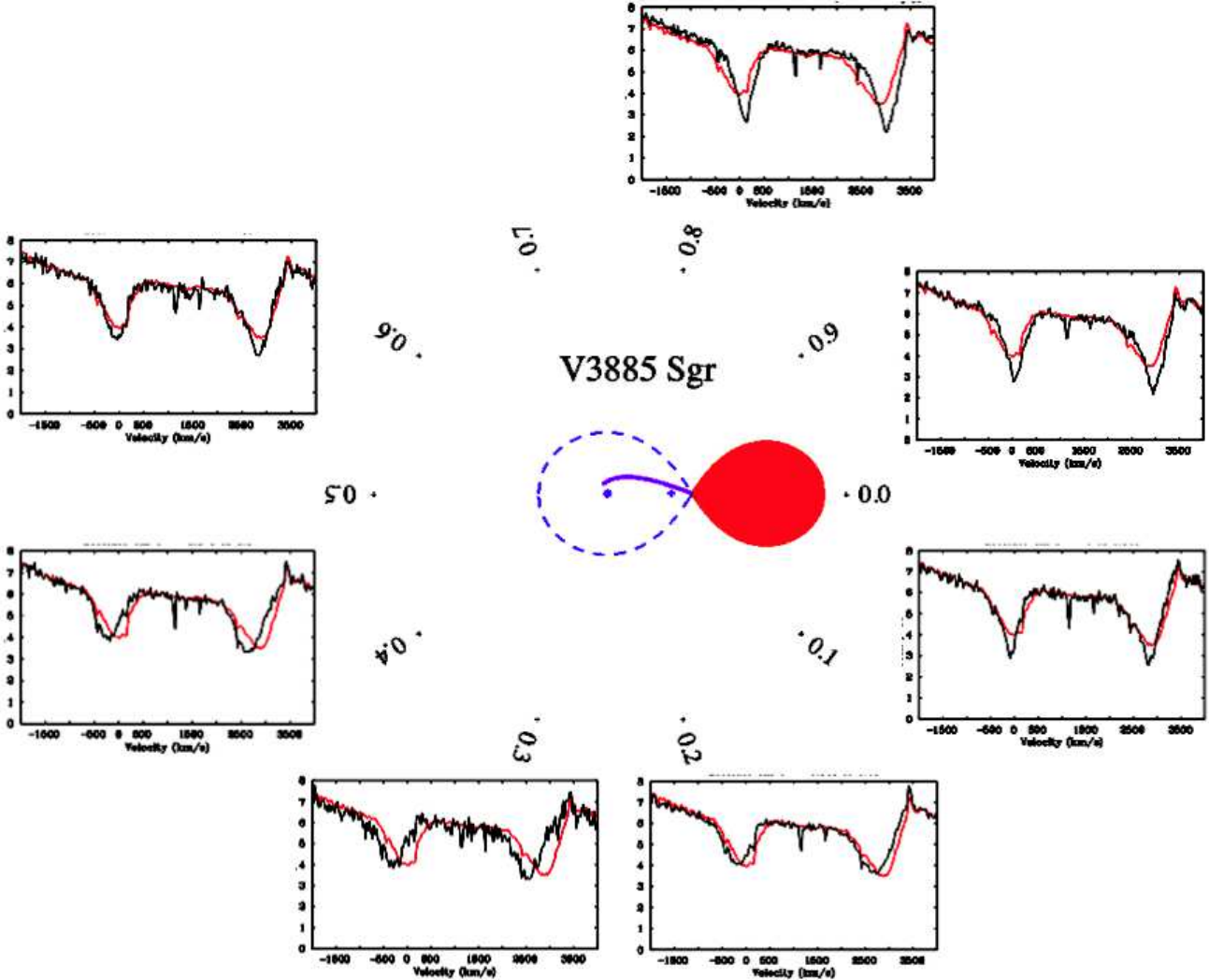}
 \caption{
Cartesian representation of V3885 Sgr showing the predicted
gravitational path of the gas stream from the donor star (centre) and the
phase-binned S{\sc iv} line profiles (black) compared to the overall mean profile
(red) for the 2000 May (Obs1) data. Phase = 0 corresponds to the positive to
negative velocity crossing of the absorption line.
}
\end{figure*}


The orbital phase dependent behaviour of the FUV lines is summarised in
Fig. 12 where we show phased-binned S{\sc iv} doublet profiles compared
to the mean profile, for the dataset of 2000 May (Obs1). The variable
line profile morphology is illustrated clearly. The absorption is
narrower and deeper at the most redward velocities (phase $\sim$ 0.75),
and the absorption is shallower when most blueshifted
(phase $\sim$ 0.25). The orbital phase-dependent changes in
line velocity and optical depth are in concert for a very wide range
of ion species, including low ions such as Ly$\beta$, C{\sc iii},
N{\sc III} and high ionization lines of P{\sc v} and O{\sc vi}.
This correspondence suggests that the FUV lines have the same physical
origin or that the material generating them is coherent enough to
supply low and high ionization regions.
The Monte Carlo based radiative transfer models of Long {\&} Knigge (2002)
predict the thermal and ionization structures of disc winds in CVs.
The disc and wind models, with plausible mass accretion rates,
provide simulations that can successfully reproduce multiple lines
in high state systems. The models can match the general cases where
C{\sc iv} and N{\sc v} are observed as well developed P~Cygni profiles,
and lower ion lines such as C{\sc iii} and N{\sc iii} appear
in absorption only.

More problematic for the line-driven disc wind model is
the origin of the axisymmetry that results in orbital modulation
of all the wind-formed lines. Time-resolved {\it FUSE} spectroscopy
of other non-eclipsing nova-like CVs has also revealed modulation
on the orbital period (e.g. RW Sex, Prinja et al. 2003;
V592 Cas, Prinja et al. 2004).
In the case of V3885 Sgr presented here it is difficult to imagine
a scenario where the orbital-phase modulations in UV lines are entirely
to due an asymmetry that has a wind-only origin.
There is no clear mechanism for shifting the velocities of the
entire bipolar outflow that arises in the
inner disc by several 100 km s$^{-1}$ in a cyclic
manner on the orbital period. 

We propose that it is more plausible that the orbital behaviour
of the FUV lines in V3885 Sgr is due to interaction(s) between
the disc and the mass accreting stream from the donor secondary.
A disc-stream interaction region, such as line emitting hot spot,
may be highlighted in the Doppler tomography presented in
Sect. 4. Also revealing in this context are the line depth
changes as a function of velocity and orbital phase (Figs. 8 and 12).
We interpret that a relatively steady component of the
blueshifted absorption lines arises from a slow (few 100 km s$^{-1}$)
disc wind in V3885 Sgr. The source of the orbital changes in the
lines may then be due to a disc-residing emission-line hot spot.
We draw some parallels here with the study of Long et al. (2009)
of the dwarf nova VW Hyi in quiescence. Long et al. propose that
significant emission in the {\it FUSE} range may arise due to
a 25,000 K hot spot where the material from the secondary encounters the
disc.
In the context of V3885 Sgr, at phase $\sim$ 0.25 the
emission source due to a hot spot has a maximum positive velocity
($\sim$ 150 km s$^{-1}$, see e.g. Fig. 8).
The superposition of this emission component on an
`underlying' disk wind absorption has the effect
of `filling in' the very low to rest velocity regions of the 
absorption trough, while the high velocity
absorption ($\simgt$ $-$200 km s$^{-1}$) is relatively unaffected.
In contrast at phase $\sim$ 0.75 the emission due to a hot spot
has maximum blueward velocity, and thus the high velocity
absorption is now affected, while the absorption
depth of the lowest velocity regions
of the profile remains comparatively high (e.g. Fig. 12).

We attempted a simple illustration of this scenario by adopting
a Gaussian profile (FWHM = 300 km s$^{-1}$; peak flux above continuum
= 1.5) to crudely represent the emission component from a localised
disc-stream interaction region, such as a hot spot.
The Gaussian profile velocity was shifted as a function of
orbital phase with a semi-amplitude of $\pm$ 150 km s$^{-1}$
(phase = 0 corresponds to red to blue crossing, and maximum redward
and blueward velocities occur at phases 0.25 and 0.75, respectively).
The product of the emission profile and the mean S{\sc iv} $\lambda$1063
profile observed in Obs1 was then used to form a short `time-series'.
The line profiles generated essentially approximate a steady
low velocity outflow with a superimposed disc residing emission source.
The phase-dependent behaviour of the emission line source is shown
in Fig. 13. Though we have simply assumed a fixed emission strength
and width (see counter evidence in Sect. 3.1),
the grey-scale image in Fig. 13 
nevertheless reproduces the
empirical result that the redward absorption between
$\sim$ 0 and 300 km s$^{-1}$ is weaker at phase $\sim$ 0.25
and more enhanced at phase $\sim$ 0.75 (cf. Figs. 7 and 8).
A more accurate reproduction of the behaviour observed in
V3885 Sgr would require ad-hoc reduction of the blueward emission line
strength at phase $\sim$ 0.75.
Ultimately it would be necessary to examine the line formation
and dynamics of a localised disc residing emission source that
in effect acts as the lower boundary to the `overlying' disc wind
in V3885 Sgr. We conclude meanwhile that the orbital phase
behaviour of FUV absorption lines in V3885 Sgr is not due
to an asymmetry in the disc wind. The line profile
variability more likely betrays a
line emission source associated with a localised region
where gas from the secondary interacts with the accretion disc.

\begin{figure}
 \includegraphics[scale=0.37]{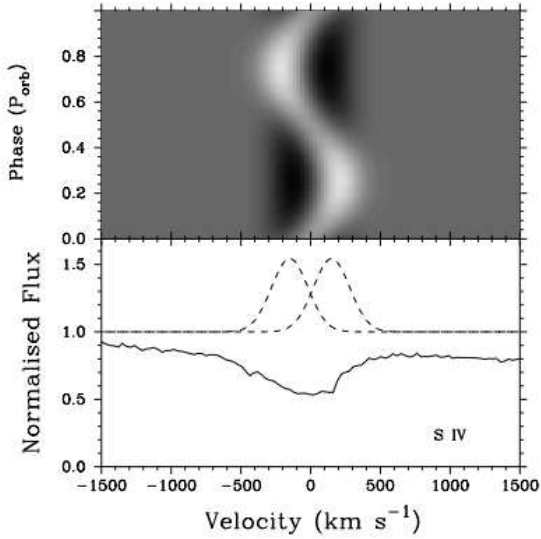}
 \caption{A simple illustration of a localised disc residing emission source
represented by a Gaussian profile (bottom panel; dotted lines)
and underlying slow disc wind. The orbital phase dependent
behaviour shown in the grey-scale image illustrates weaker
absorption redward of rest velocity at phase $\sim$ 0.25 compared to
phase $\sim$ 0.75.}
\end{figure}

\section{Concluding summary}

We have analysed the variability characteristics of {\it FUSE}
(TTAG) time-series spectra of the non-eclipsing, nova-like
system V3885 Sgr. The data were secured over two epochs in
2000 and 2003. Our key results are:

\noindent
(i) The low (C{\sc iii}), intermediate (S{\sc iv}) and high
(P{\sc v}, O{\sc vi}) ionization lines vary in a tightly
corresponding manner, on time-scales of a fraction of
the system orbital period ($\sim$ 0.21 days).

\noindent
(ii) The line profile variability is predominantly orbital-phase
modulated, with a S-wave velocity motion of the observed absorption
troughs over a range of $\sim$ $-$200 km s$^{-1}$ to +200 km s$^{-1}$.

\noindent
(iii) For phase = 0 corresponding to the +ve to $-$ve crossing
of the spectral lines, the maximum blueward velocity occurs at phase $\sim$ 0.25
and the maximum redward displacement is at $\sim$ 0.75.

\noindent
(iv) In addition to the orbital modulations, stochastic line profile
changes are evident on time-scale down to $\sim$ 3 minutes.
We associate these events with instabilities in the accretion stream
or disc wind.

\noindent
(v) We have derived the first FUV tomograms of V3385 Sgr,
to complement previously published tomography based on optical
spectra.
With the caveat that the FUV lines likely arise from multiple
line formation regions, we find evidence in the tomograms
for structures forming where the gas stream interacts with the
accretion disc.

\noindent
(vi) We favour a scenario where localised disc-stream interaction
generates a line-emitting source, and thus provides the dominant
axisymmetry for orbital modulated FUV line profile behaviour
in V3885 Sgr.

\section*{Acknowledgements}
This research was partially supported by NSF grant AST-0908440 to MTR.
LWP acknowledges support from STFC studentship.
We thank an anonymous referee for helpful suggestions that
improved this paper.

\bsp

\label{lastpage}

\end{document}